\documentclass{llncs}
\usepackage{xspace}
\usepackage[latin1]{inputenc} 
\usepackage[T1]{fontenc}
\usepackage{textcomp}

\newif\ifpdf
\ifx\pdfoutput\undefined
 \pdffalse % we are not running PDFLaTeX
\else
 \pdfoutput=1 % we are running PDFLaTeX
 \pdftrue
\fi 
\ifpdf % stuff wanted only when using PDFTeX and hyperref
  \usepackage[pdftex]{graphicx} % graphicx with PDFTeX driver to include BMPs, JPGs, ...
  \usepackage{hyperref}  % no driver for PDFTeX, otherwise use dvips
   \hypersetup{%
     pdftitle= {Transitive trust in mobile scenarios},
     pdfauthor= {Nicolai Kuntze, Andreas U. Schmidt},
     pdfkeywords= {Trusted computing, transitive trust, authentication,
service access, mobile network} 
}
\else
  \usepackage[dvips]{graphicx} % um .eps Dateien einzubinden 
  \usepackage[dvips,pagebackref]{hyperref}
\fi
\begin{document}
\mainmatter              % start of the contributions
\pagestyle{plain}
\title{Transitive trust in mobile scenarios}
\titlerunning{Transitive trust in mobile scenarios}  
%                                     abbreviated title (for running head)
%                                     \toctitle is used%
\author{Nicolai Kuntze, Andreas U. Schmidt}
\authorrunning{N. Kuntze \&  A. U. Schmidt}   % abbreviated author list (for running head)
%
%%%% modified list of authors for the TOC (add the affiliations)
\tocauthor{Nicolai Kuntze, Andreas U. Schmidt (Fraunhofer SIT)}
\institute{Fraunhofer Institute for Secure Information Technology SIT,\\ 
Rheinstrasse 75, 64295 Darmstadt, Germany,\\
\email{\{Nicolai.Kuntze,Andreas.U.Schmidt\}@sit.fraunhofer.de},\\ WWW home page:
\href{http://www.sit.fraunhofer.de}{www.sit.fraunhofer.de},
\href{http://www.math.uni-frankfurt.de/~aschmidt}{www.math.uni-frankfurt.de/\~{}aschmidt}}

\maketitle              % typeset the title of the contribution
\setcounter{footnote}{0}
\begin{abstract}
Horizontal integration of access technologies to networks and services
should be accompanied by some kind of convergence 
of authentication technologies. 
The missing link for the federation of user identities
across the technological boundaries separating authentication
methods can be provided by trusted computing platforms.
The concept of establishing transitive trust by trusted computing
enables the desired cross-domain authentication functionality.
The focus of target application scenarios lies in the realm of 
mobile networks and devices.
\end{abstract}
% custom macros
%
\section{Introduction}
Current information technology imposes on users a multitude of heterogeneous 
authentication mechanisms when they want to access networks, services, or content.
The technical access channels to these desiderata are, however, undergoing a
continual process of convergence.
The mobile domain provides a striking example~\cite{MAUS05}.
The access to services through mobile devices shows a trend to 
become network-agnostic. 
Driven by the horizontal integration of technologies, 
users will soon be able to consume services seamlessly from a single device 
via a variety of channels and transport methods such as 
2G, 3G, WLAN, Bluetooth, WiMAX, MobileIP, or the upcoming Zigbee. 
Accordingly, end users' attention will shift away from the pricing of bandwidth
to that of content and services. 
Custom must then be attracted by offering applications and content with good 
price to quality ratio.
Little room is left for returns generated by charging for network access and data transport.
Business models necessarily undergo drastic changes, 
of which the mushrooming of virtual network operators
is the salient epiphenomenon.
Research has long forseen this evolution toward `value networks'~\cite{LW02,ULS02}.

Thus, information networks are becoming ever more service oriented.
On the application layer, identity management (IDM), 
as embodied, e.g., in the Liberty alliance standard suite, has proved to be
a successful foundation for the user-centric integration of service access~\cite{CK01}.
Mobile networks with millions of users and even more identities are already 
using IDM for essential services like roaming~\cite{RAN04}.
Yet, arguably, these top-level methods require infrastructural 
support of some kind~\cite{LOP04}.
In particular, it is desirable to overcome the boundaries between logically,
technically, or even physically separated domains and their respective authentication
methods.
This signifies a second layer of technological convergence, namely convergence of 
authentication methods and the domains of trust defined by them.
This is the subject matter of the present paper.

We argue that trusted computing (TC) can be a means to the above mentioned ends.
In fact, two systems or devices can assure each other of their being in a 
trustworthy state through TC methods like direct attestation.
If the devices carry credentials from various trust domains, they can 
then use TC-secured communication to exchange them.
This assignment of credentials by trustworthy transmission between
carriers yields \textit{transitive trust relationships}.
This allows for the mediation of trust between domains and
user or device identities, and in fact, some of the concepts 
we present are rather similar to logical identity federation.
However, transitive trust by TC enables the traversal of authentication domains
hitherto separated by technical or even physical boundaries.
The concept of transitivity of trust relationships was recently
analysed in~\cite{JGK06}. 

The paper is organised as follows. Section~\ref{sec:TT} explains the
basic notions behind transitive trust, in particular the three most 
primitive operations supported by it. The exposition, while theoretical,
is not completely formalised in view of the intended application scenarios. 
Three of the latter scenarios are described in ascending level of
detail in Section~\ref{sec:SCEN}.

Not by coincidence are these applications chosen from the mobile realm.
In fact we show that mobile devices equipped with TC are not only good 
carriers for credentials but also excellent links between trust domains,
when applying the methods of transitive trust.
As will become clear from the few scenarios we consider, potential 
business models, enabled by transitive trust, abound.
Needless to say, the newly conceived trust relationships that 
we describe in concrete business scenarios must be supported in the
real world by contractual relationships.
\section{Transitive trust by trusted platforms}
\label{sec:TT}
A completely formalised definition is outside of the scope of the 
present paper, since we aim at rather specific application scenarios.
Nevertheless we want to provide a theoretical descriptions that
allows to assess the generic character of the transitive trust
relationships supported by trusted platforms, i.e., systems
secured by TC as described below.
A more formal treatment, e.g., along the lines of~\cite{JGK06} or~\cite{MAU96},
is certainly possible. Yet, it would not contribute much to the present topic
since we are more interested in pinpointing the properties and functionalities
of trusted platforms involved in the establishment of transitive trust. 

We use a simple model for actors in trust domains consisting of 
trust \textit{principals} and \textit{agents}.
Trust principals are the subjects defining an authentication domain
by issuing credentials to users or enrolling them to their devices.
They control domain membership and applicable authentication methods,
and therefore define a domain of trust like an identity provider.
Trust principals are denoted by capital letters $A$, $B$, $C,\ldots$.
Agents asking for access to services provided in a certain domain
are denoted by  $a$, $b$, $c,\ldots$. The notion of agent signifies 
\textit{classes} of individuals, i.e., groups of agents
who enjoy the same access rights in a certain application 
context when authenticated using their respective (individual) 
credentials. A subgroup of agents is written as $a'\subset a$ as usual.

Credentials $\gamma_{a,A}$ are objects or data which authenticate agents
$a$ with respect to a principal $A$. We do not specify the particular
kind of credentials used, nor the accompanying authentication methods.
This notion is very generic and comprises classical examples like
SIM/USIM, Hardware tokens, Smartcards, PKI-based certificates,
PIN/TAN-based methods, or even personal credentials, e.g., 
Machine Readable Transfer Documents or a health (professional) card. %(MRTD)

It should be clear that the overall security of the authentication 
assertions of transitive trust that are described below depend on 
the 'weakest link' in the trust chain. These assertions can in particular 
not be stronger than those provided by the original credentials.
Furthermore, the trust scope implicated by a successful
authentication, i.e., the specific type 
of trust assumed in a given principal-agent relationship, may vary
from domain to domain.
As already mentioned, risks arising from these complexities must be
assessed and mitigated in the context of the specific application
scenario at hand.
Common instruments for that are contracts between principals
and their agents and bridging contracts between principals.
\subsection{Trust credentials}
Credentials that can be constructed basing 
on the functionalities of a trusted platform
module (TPM~\cite{TPM1285}) play a special role in our concept.
TPMs provide a number of features that can be used to securely
operate a system.
Methods for the secure generation, 
storage, and usage of asymmetric key pairs are the foundation 
for encrypted and authenticated operation and communication.
Trust measurements on the system environment 
exerted at boot- and run-time allow for trustworthy assertions
about the current system state and a re-tracing of how it was reached.
The system state is securely stored in platform configuration 
registers (PCR) tamper-resistantly located inside the TPM.
Memory curtaining and sealed storage spaces are enabled by
pertinent TPM base functions.
Trustworthy system and application software can build on this basis
to establish authenticated communication with the exterior and
transmit data maintaining integrity and confidentiality.
In particular, Direct Anonymous Attestation (DAA), a method put 
forward in~\cite{BCC04} and specified by the TCG, enables 
the establishment of trust relationships
of a trusted system with external entities.
A central goal of DAA is to cover privacy issues related
to previous versions of the standards~\cite{CAM04}.

Although certain flaws are known in the TCG standards
(e.g.~\cite{BCLM05} points to a flaw in the
OIA Protocol an authorisation protocol which represents one of
the building blocks of the TPM)
that exist currently
future versions are likely to remedy them.
We assume for the purport of our applications that
the functions used are at least 
secured against common attack vectors
in the scenarios below.

%privacy issues, DAA

Using the described functionality, a trusted system,
viewed as an agent $a$, can establish what we call a 
\textit{trust credential} $\tau_a$. 
Specifically, we assume that the trust credential can
be used to attest the validity of 
three fundamental security assertions
of a system to the exterior.
\begin{enumerate}
\item The presence of a live and unaltered TPM. 
This can for instance be carried out using a 
challenge-response method using the TPM's endorsement 
credential. Endorsement credentials are pre-installed by the TPM's 
manufacturer.
\item The integrity of the system and its components. 
This property is ascertained through trust measurements and
communicated via DAA.
\item That an existing credential $\gamma_{a,A}$ is unaltered.
This must be established by trusted system software and components
used to access the credential's data. Again, this assertion is 
forwarded to other parties using direct attestation and 
secure communication channels established therewith.
\end{enumerate}
These properties are not independent but build on each other, i.e,
to prove 3.\ one needs first attestation of 2.\ and 1., etc.
The TPM is capable of creating, managing, and transmitting own
cryptographic credentials which can 
convey the described assertions 1.--3.

We now describe three basic, independent 
operations for creating trust between
agents and principals. 
These methods represent the essence
of transitive trust enabled by trusted platforms.
They all rely on \textit{referral trust} in the parlance of~\cite{JGK06}.
That is, on the ability
of a trusted agent through assertions 1.--3., to 
make recommendations to trust another agent or even himself in
a special, functional role.

%Another property can be attested, on the same level as 3.
%\begin{enumerate}
%\setcounter{enumi}{3}
%\item The integrity of additional credentials that 
%can be securely enrolled by some 
%principal at any time.
%\end{enumerate}
%
\subsection{Restriction}
\label{sec:res}
By the method of restriction, a subgroup of agents
$a'\subset a$ belonging to the authentication domain of principal
$A$ can be defined.
Agents of class $a$ authenticate themselves in the conventional
way associated to their credential $\gamma_{a,A}$.
This establishes an authenticated channel, over which 
agents of subclass $a'$ transmit an additional trust
credential $\tau_{a'}$ identifying them as members of $a'$.
Since by this method the trust and original credentials are used
independently, only assertions 1.\ and 2.\ are needed.

The additional security and in effect higher trust in
agents of $a'$ provided by them allows to ascribe
to $a'$ more service access rights than to $a$-agents.
In particular, the integrity of client software 
can be attested by 2.
Those clients can access content
or services only available to the privileged subgroup.
This is in fact the classical scenario used to enforce
copyright protection through digital rights management (DRM).
A higher security level is provided by restriction in a
very generic way. The possibility for $A$ to check the 
consistency of the trust credential $\tau_{a'}$ with that of
$\gamma_{a,A}$ makes at least the subclass $a'$ more resilient
against cloning attacks on the credential $\gamma_{a,A}$.
This kind of attack is not uncommon in the mobile sector~\cite{CHE05}.

This raised resilience against cloning is the main reason
why the usage of a trust credential is advantageous 
for the definition of the subclass $a'$. The latter definition 
can be implemented in various ways.
The first-best approach is restriction under 
the authority of the principal.
She can manage access control lists based on
\textit{individual} trust credentials 
identifying a single TPM.
Or, e.g., she can use individual trust credentials to establish
a secure channel with $a'$-agents and distribute a shared
secret to them. This secret can reside in the part 
of the system protected by the TPM and thus become part
of $\tau_{a'}$. In turn it may be used in subsequent authentication
requests toward $A$, keeping an agent's individual identity secret.

A proper choice of enrolment method and time for the trust credential
is essential for the validity of the additional trust
provided by the restriction operation. 
If the credentials $\gamma$ and $\tau$ are impressed on the agents
independently of each other, i.e., not both under the control of
the principal $A$, then, e.g., resilience against cloning attacks
is restricted.
Since $A$ cannot associate the two credentials belonging to an
individual agent, she can at best 
avoid to grant two agents with identical $\gamma$ service access
by using a first-come-first-served approach.
It is possible to improve on this by forcing an activation
of $\tau_{a'}$ at an early stage, e.g., the time of roll-out 
of a mobile device.
Higher cloning-resilience can only be achieved if the principal
individualises both credentials and controls their deployment
to the agent.

It may be more the rule than the exception that the trust credential
$\tau_{a'}$ provides stronger authentication than the original one
$\gamma_{a'\subset a,A}$. Conventionally, $\tau$ would then be the preferable
credential to authenticate agents of class $a'$ with.
It is essential for the understanding of the present concepts to 
notice that this is often not practical. Namely, the communication
channel through which $\tau$ is conveyed to the principal is only available
after authentication by $\gamma$. A paradigm is the access
to mobile networks as described in section~\ref{sec:prepaid}.
\subsection{Subordination}
\label{sec:sub}
By subordination an agent $a$ in principal $A$'s domain can enable
the access to this domain, or certain services of it, for another
agent $a'$. By this, $a'$ is effectively included in $A$'s domain of trust,
respectively, $A$'s domain is extended to $a'$.
As for restriction, $a$ authenticates himself using a generic credential
$\gamma_{a,A}$ and then produces a specific trust credential $\sigma_{a}$
identifying those agents of $A$'s domain who are allowed to dominate
certain other agents.
The subordinated agent $a'$ shows a trust credential $\sigma_{a'}$ to $a$,
who in turn mediates the access to $A$'s services, either
by forwarding authorisation requests, or granting them himself.
Furthermore, the authentication of $a$ and $a'$ can also be mutual
rather than one-sided.

Implementation variants of this operation and authorisation
based on it are manifold, despite its simplicity.
The most restrictive approach would be to use the secure communication
channels between $a$ and $a'$ 
(mutually authenticated by $\sigma_{a}$, $\sigma_{a'}$), 
and $a$ and $A$ to forward every single authorisation
request from $a'$ to $A$ including the trust credential $\sigma_{a'}$.
Independently of the degree to which $A$ takes part in authorisation,
the act of authentication for subordination is generically between 
$a'$ and $a$. 
Nevertheless, in many scenarios $\sigma_{a'}$ is controlled and
enrolled by $A$, and the principal can in implementation
variants also partake in authentication, e.g., by facilitating
steps in a challenge-response protocol.

If genuine trust credentials are used for subordination, 
the operation employs only TPM functions 1.~and 2.~above.
TPMs provide user functions for the revocation of keys,
which is a point of failure in this case.
Thus one might use a dedicated credential
$\gamma_{a',A}$ for subordination.
Such a credential should then live in the trusted part of the
subordinated system and be secured in the authentication by
function 3.~to mitigate forgery.

A subordination scenario is outlined in~\ref{sec:bond}.
\subsection{Transposition}
\label{sec:trans}
Transposition operates between the trust domains of two principals
$A$ and $B$.
The authentication of an agent $b$ of $B$'s domain is mediated
by an agent $a$ of $A$'s domain and the principal $A$.
This can make sense for instance if direct communication between
$b$ and $B$ is not possible as in the scenario of Section~\ref{sec:Pos}.

We assume that authentication of $a$ to $A$ is done as above.
Trust credentials $\tau_a$ and $\tau_b$ are used for (mutual) authentication
of $b$ to $a$ (or between them).
Here, the third TC function of $\tau_b$ is used to prove the integrity
of a credential $\gamma_{b,B}$ with which $b$ is ultimately authenticated
with respect to $B$.
The generic situation for the latter authentication is as follows.
The credential $\gamma_{b,B}$ is forwarded to $A$. 
This bears the assurance that an authentic (by $\gamma_{a,A}$) and 
untampered (by $\tau_a$) agent has handled the latter credential.
In effect $a$ establishes a trusted path for the transmission
of $\gamma_{b,B}$. 
Whether or how $\gamma_{b,B}$ is transferred from $A$ to $B$ to
finally authenticate $b$ depends on communication means and 
contractual relations.
The transposition concept leaves this open.

Again, transposition can be implemented in numerous variants.
In particular, part or all of the functionality necessary
for authentication of $b$ can be deferred to $A$ or $a$.
From $B$'s perspective, efficiency gains by such an outsourcing
or even decentralised approach to authentication must be balanced
with the protection of secrecy of his business data and processes,
which, to a certain extent have to be turned over to $A$.

On the other hand, in the generic transposition operation where $\gamma_{b,B}$ is 
forwarded to $B$ who in turn completely controls the authentication of
$b$. Then, additional cryptographic means can be applied to render
any sensitive information about the relationship of $b$ and $B$
inaccessible to $a$ and $A$.
In particular, $B$ might want to keep his agents anonymous to $A$,
and even the mere size of $B$'s domain of trust might be an informational 
asset worth of protection.
\section{Scenarios}
\label{sec:SCEN}
This section outlines three concrete application scenarios
of economical relevance, corresponding to the three operations
explained above. The first two are sketched on a rather high level,
while the third and most complex one is used to detail 
processes and protocols. A detailed description of the first
two scenarios would be very similar.
\subsection{Functional discrimination of mobile devices}
\label{sec:prepaid}
As already said, the paradigm for restriction scenarios is DRM.
We want to pursue a slightly different direction and take a look at
the relationship between network operator and customer in the mobile 
domain. 
The standard form of customer retention exerted by the MNO
is SIM-lock, a crude form of functional restriction of mobile devices
bonding mobile devices to SIMs of a certain MNO.
Based on transitive trust restriction, a finer grained functional
discrimination of mobile devices becomes possible.
Depending on the device vendor's and MNO's business models, various
client functions of the device can be restricted to certain, 
more or less privileged customer groups.
The management of mobile devices, of which functional
discrimination is an important instance is viewed by the industry
as a fundamental application area of TC~\cite{TMP}.

A multitude of benefits accrue to MNO and customer in this kind of scenario.
First, it is cost-efficient to produce a single product line with many 
appearances to the end-user, rather than marketing a multitude of makes 
and models as customary today.
Second, the up- and downgrading of functionalities can be 
implemented dynamically, without physical access to the device.
To the user, the relative seamlessness with which device 
control operates is an ergonomic benefit and allows for better
customisation and even personalisation.

The efficient means to implement functional restrictions of mobile
devices is provided by the trusted boot process and operating system of the 
trusted platform it represents.
Thereby, the trust credential can attest two properties via DAA.
First, that the device belongs to a certain,
restricted group defined explicitly or implicitly by a list of enabled functions.
Second, that the device actually is in a state where only the allowed
functions are enabled.
The set of functions to be managed could be pre-configured
and the dynamic control effected via simple changes of parameters,
e.g., for values in PCRs.

The enforcement level of this approach is stronger as compared to
SIM-lock precisely because the trusted platform's base operation
software is tamper resistant.
Based on this assurance, the MNO can deliver specific services or
content only to the restricted group privy to it. 
Thus functional restriction provides the foundation on the
client side for further service discrimination, policy enforcement,
and DRM proper.

As a simple instance using the 
transitive trust restriction operation, a prepaid mobile phone
can be implemented.
The phone carries  in its trusted storage area a running total
which is decremented by a trusted software.
While the initial access to the mobile network is still established
using SIM authentication, DAA and the trust credential
then yield assurance to the MNO that the running total is nonzero,
upon which access to the network's communication services can be granted.
This releases the MNO from operating (or paying for) a centralised accounting.
\subsection{Bonding of mobile accessories} %Bonding of mobile accessories
\label{sec:bond}
For the mobile domain, an application of subordination which suggests 
itself is to extend the authentication of devices toward an MNO to
devices not equipped with SIM cards or even physical access to
the mobile network.
A commercial application is the extension of SIM-lock to such devices.
For the purpose of customer retention, such a scheme can for instance be
combined with loyalty programmes.
Just as SIM-locked mobile phones are highly subsidised, an MNO can give
away technical accessories such as digital cameras, media players, or 
high quality headsets.
The functioning of those subordinated devices is then 
dependent on authentication toward a mobile device 
or any device in a specific MNO's network.

In effect, the accessories can be given away for a very low price
or even for free on the condition that they work only within
the subsidising  MNO's network.
The devices are bonded to the MNO.
As an additional benefit for the MNO, the traffic generated by subordinated
devices is bound to his own network (as traffic volume is
a traditional economic value indicator for MNO businesses).
Of course, advanced service provisioning can be based on accessory bonding,
e.g., the MNO or another provider can offer storage, organisation,
and printing services for photographs taken with a bonded camera.
\subsection{Point of sales}
\label{sec:Pos}
We now come to scenarios employing the transposition operation, and here present the
related technical processes and communication protocols in some detail.

A user with a TPM-equipped  mobile device wants to purchase a soft drink from a 
likewise trust-enabled vending machine, the point of sales (POS). 
While the user still makes up her mind on her taste preferences, 
device and POS initiate a trusted communication session using DAA 
and transport layer encryption. 
Device and POS thus achieve mutual assurance that they are in an unaltered, 
trustworthy state, and begin to exchange price lists and payment modalities. 
After the user selects a good and confirms his choice at his device, 
signed price and payment processing information is transferred to the MNO. 
After verifying the signatures and optionally informing the good's vendor 
and a payment service provider, the MNO sends a signed acknowledgement 
to the mobile device, which relays it to the POS, where it is verified 
and the good is delivered. 

The benefits for the vendor that arise in this scenario basically stem 
from the transitive trust relationship that is mediated between MNO and 
POS by the mobile device. 
It entails in particular that no network communication is required during 
the initiation of a trusted session, that no transaction data needs to be 
stored in the POS, and that, ultimately, the POS does not need to be equipped 
with networking capabilities --- at least for the sales process. 
In this way the MNO provides payment services as well as authorisation 
control for the vendor. 
This requires little more than a TPM and a short-range communication module 
in the vending machine.
In extended service scenarios, the customer's mobile devices can as well
be utilised to transfer valuable information to the POS, e.g., updated price
and commodity lists, or firmware. 

A similar example regards home 
automation and lets a user and her mobile device become part of the 
maintenance service of, say, the heating system of her home. 
Based again on their respective TPMs, heating system and mobile 
device establish a secure communication channel to exchange maintenance 
data, or data used for metering. 
This can be done both at specific user requests or even seamlessly 
during normal operation of device and heating system, 
every time the machine-to-machine communication 
module of the device gets in the range of the one in the heating system. 
In this way, the mobile device can notify user and a maintenance 
chain about necessary repairs and also support accounting and billing. 
Here, a trusted computing approach not only ensures the protection of 
personal data, it also
enables a simple means of remote maintenance and home automation 
in non-networked homes by efficiently utilising the mobile network.

\begin{figure}[t!]
\centering \ifpdf \resizebox{1.0\textwidth}{!}{\includegraphics{diagram.pdf}} \else
\resizebox{1.0\textwidth}{!}{\includegraphics{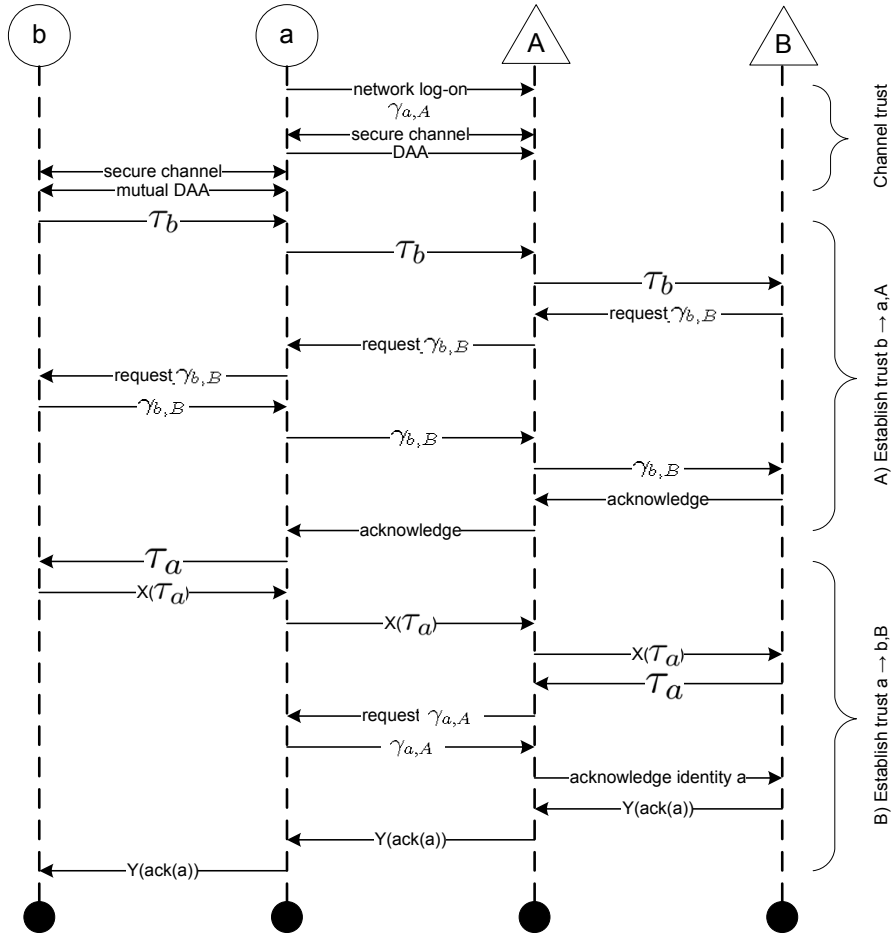}}
\fi
  \caption{Sequence diagram for the transposition operation from POS $b$ 
via mobile device $a$, MNO $A$, to POS owner $B$. The notation $X(\cdot)$, $Y(\cdot)$
means protection by secrets $X$, $Y$ shared between $b$ and $B$.}
  \label{diagram}
\end{figure}
Returning to the POS scenario, we now describe one possible implementation
in more detail. We concentrate on the authentication processes and leave
selection, purchase, and payment aside.

The variant of transposition we consider is that of 
\textit{maximal mutual trust}. 
That is, both principals $A$, the MNO, and
$B$, the POS' owner, can trust the involved agent of the other
domian, i.e., the POS $b$, respectively the mobile device $a$.
The raised level of security ensuing from this may be desirable in 
particular from $B$'s perspective, depending on the sensitivity 
of business data handled by $a$ and $A$ as mediators, for instance
if accounting and charging services of $B$ are transferred to $A$.
The process to achieve this kind of 
transposition can be divided into 
two principally independent steps.
\begin{enumerate}
\item[A)] Establishment of trust of $a$ and $A$ in agent $b$. 
\item[B)] Establishment of trust of $b$ and $B$ in agent $a$. 
\end{enumerate}
These two steps are in fact equivalent to two subordination
operations with exchanged roles.
A sequence diagram for both steps is shown in Figure~\ref{diagram}.
Note that A) and B) can be interchanged or even overlap.

The two main steps must both be preceded by an establishment
of a secure communication channel between $b$ and $a$ and between
$a$ and $A$, respectively.
For the latter, the usual log-on of the mobile device to the network
based on $\gamma_{a,A}$ is augmented by  attestation of the trusted platform
$a$ via DAA toward $A$ over a secured channel based on, say, encryption 
on the transport layer.
For the former, mutual platform attestation over an encrypted channel
is carried out between $b$ and $a$.

\begin{enumerate}
\item[A)]
The trust credential $\tau_b$ of b is passed on to $B$, attesting to
$B$ that there is one of his  untampered POS down the communication line.
$B$ then requests and receives proper authentication from $b$ with $\gamma_{b.B}$.
The underlying assumption that $B$ can associate trust and generic credentials of
agents in his domain is  a central anchor for trust in the present variant 
of transposition. 
In effect $B$ is an identity provider for trust credentials of his domain.

$B$ acknowledges successful authentication of $b$ to $A$ who passes it on
to $a$. The trust relationship between the two principals and
$A$ and his agent $a$ assures the latter two actor of the authenticity of $b$.

\item[B)]
Agent $a$ initiates his authentication toward $B$ and $b$ by handing his
trust credential to $b$.
This credential cannot be utilised by $b$ directly to authenticate $a$, but is 
rather used as a pledge which is then redeemed by $b$ at the principals.
To that end, $b$ uses some secret $X$ he shares with his principal to
protect $\tau_a$. 
$X$ can for instance be established using the Diffie-Hellman method~\cite{DH76}. 
The protection of $\tau_a$ by $X$ prevents $a$ and $A$ from tampering with
the authentication request that is embodied in the message $X(\tau_a)$ passed
on to $B$.

It should be noted that, apart from transport and addressing information,
$a$ and $A$ need not know for which of $A$'s agents authentication is 
requested, if $X$ comprises encryption. 
Thus, the identity of the authenticated agent $a$ 
could be kept secret from $A$ in an advanced scenario.
This could be used to protect the privacy of agents in the domain
of $A$, e.g., with respect to their purchasing patterns.

$B$ sends $\tau_a$ to $A$ and with that requests from $A$ the authentication of it.
If $A$ does not have a registry of all valid trust credentials in his domain
or any other means of authenticating them then $A$ has to exert a secondary
authentication of $a$ by the generic credential $\gamma_{a,A}$ (again assuming
that association of $\tau_a$ to $\gamma_a$ is possible).
$A$ acknowledges the identity of $a$ to $B$.
This acknowledgement is passed on from $B$ to $b$, again protected by a shared
secret $Y$ to prevent tampering with it on its way.
\end{enumerate}
\section{Conclusions}
\label{sec:CONC}
We introduced the notion of transitive trust for a pragmatic purport.
It is intended as a conceptual blueprint for the systematic construction
of concrete, TC-based application scenarios. The examples exhibited
show that transitive trust has a potential to be a fertile
concept to that end.
In particular, new application and 
business scenarios are enabled by transitive trust
as well as more efficient and/or more secure implementations of old ones.
Protection of privacy is not in opposition to the use of TC in those
scenarios.
It can, on the contrary, be supported in carefully constructed 
implementation variants of transitive trust.

As said, transitive trust is very similar to (a subset of) 
identity federation.
Economically the prospect to federate the identities of millions of
subscribers of mobile networks with other providers of goods and services, 
is rather attractive.
TC has additional application potential due to the possibility 
to transgress boundaries of authentication domains that are closed to IDM
on the application layer.

A particular trait of transitive trust mentioned above is the 
enabling of de-centralised authentication through the trusted agents.
A benefit of such approaches can be enhanced resilience and availability
of service access.
They can also be a base for de-centralised authorisation and ultimately
de-centralised business models, such as super-distribution of virtual 
goods from agent to agent, cf.~\cite{AUSAX05,SEAX05,RAJAX05}.

As a further example, in an advanced scenario for the restriction operation, it can
be envisaged that a group of agents defines itself in a manner
similar to building a web of trust~\cite{KR97} of which PGP is a well-known
instance~\cite{ZIM}.
To that end, the transposition operation could be used to establish mutual trust
between agents, extend it to trust paths in a community, and eventually
define the subgroup as the resulting web of trust.

%
\end{document}